\documentclass[3p,times,twocolumn]{elsarticle}

\usepackage{amsmath,amssymb}
\usepackage{dcolumn}
\usepackage{bm}
\usepackage{tikz}
\usepackage{pgfplots}
\usepackage{float}
\usepackage{subcaption}
\usepackage{ragged2e}
\usepackage{helvet}
\usepackage[eulergreek]{sansmath}
\usepackage{hyperref}

\pgfplotsset{compat=1.18,
    xlabel shift=-2pt,
    ylabel shift=-4pt,
    every picture/.append style={trim axis left, trim axis right},
    every axis label/.append style={font=\sansmath\sffamily\footnotesize},
    every axis title/.append style={font=\sansmath\sffamily\footnotesize, yshift=-0.75ex},
    every tick label/.append style={font=\sansmath\sffamily\footnotesize},
    every axis legend/.append style={font=\sansmath\sffamily\footnotesize}
}
\usepgfplotslibrary{fillbetween}

\makeatletter
\g@addto@macro\bfseries{\boldmath}
\makeatother

\definecolor{forestgreen}{rgb}{0.0, 0.27, 0.13}
\definecolor{darkblue}{rgb}{0.0, 0.0, 0.55}
\definecolor{darkred}{rgb}{0.55, 0.0, 0.0}
\hypersetup{colorlinks,
           linkcolor={forestgreen},
           citecolor={darkblue},
           urlcolor={darkred},
           pdftitle={Neural Networks Reveal a Universal Bias in Conformal Correlators},
           pdfdisplaydoctitle
}

\def\GG{{\cal G}}
\def\HH{{\cal H}}

\def\MM{{\cal M}}
\def\NN{{\cal N}}
\def\OO{{\cal O}}

\def\IR{{\mathbb R}}

\def\bal{\begin{align}}
\def\eal{\end{align}}

\def\Df{{\Delta_\phi}}

\makeatletter
\newcommand{\customlabel}[2]{%
   \protected@edef\@currentlabel{#2}%
   \label{#1}%
}
\makeatother

\journal{Physics Letters B}

\begin{document}

% ltex: enabled=true

\begin{frontmatter}

\title{Neural Networks Reveal a Universal Bias in Conformal Correlators\tnoteref{t1}}
\tnotetext[t1]{Preprints CCTP-2026-4, ITCP-2026-4}

\author[kcl]{Kausik Ghosh}
\ead{kau.rock91@gmail.com}

\author[kcl]{Sidhaarth Kumar}
\ead{sidhaarth.kumar@kcl.ac.uk}

\author[crete]{Vasilis Niarchos}
\ead{niarchos@physics.uoc.gr}

\author[kcl]{Andreas Stergiou}
\ead{andreas.stergiou@kcl.ac.uk}

\address[kcl]{Department of Mathematics, King's College London, Strand, London WC2R 2LS, United Kingdom}

\address[crete]{Department of Physics, CCTP and ITCP, University of Crete, 71303, Greece}

\begin{abstract}
We propose that simple neural networks (NNs) trained on crossing symmetry can reconstruct conformal correlators restricted to a line to remarkable accuracy. The input is minimal: an external scaling dimension, a spectral gap, and the value of the correlator at a single point. We present evidence across a wide range of conformal theories and dimensions, for both four-point and thermal two-point functions. We attribute these observations to the \textit{spectral bias} of gradient-based NN training, which appears to align with an intrinsic smoothness property of conformal field theory. This suggests a novel variational principle for conformal correlators and opens a path towards a powerful new computational framework for non-perturbative quantum field theory.
\end{abstract}


\end{frontmatter}

\section{Introduction}
\label{intro}

The solution of interacting quantum field theories (QFTs) remains to-date a challenging central problem in many physical applications. Since typical QFTs can be viewed as renormalisation group flows between different conformal field theories (CFTs), a truly non-perturbative treatment of CFTs would be a significant step towards a deeper understanding of quantum fields in general. %, as well as quantum gravity.

Consider a CFT in flat space. What does it mean to solve that CFT? For local physics, a solution should determine the spectrum of quantum operators and their (multi-local) correlation functions. The quantum operators are labelled by their scaling dimensions $\Delta$, their spin $J$ and possibly other global symmetry quantum numbers. The correlation functions are multi-variable functions of the spacetime coordinates. The computation of most of these data is a daunting task, typically beyond the reach of existing analytic and numerical techniques.

In the modern conformal bootstrap approach, symmetries and mathematical consistency are leveraged to extract information about the underlying CFT data. For example, conformal symmetry implies that the four-point function of a scalar operator $\phi$, with scaling dimension $\Delta_\phi$, takes the form
\begin{equation}
    \label{introaa}
    \langle \phi(x_1) \phi(x_2) \phi(x_3) \phi(x_4) \rangle = \frac{1}{(x_{12}^2 x_{34}^2)^{\Delta_\phi}}\,\mathcal{G}(z,\bar z)\,,
\end{equation}
where $x_{ij}^2 = (x_i-x_j)^2$, and $z\bar z=\frac{x_{12}^2 x_{34}^2}{x_{13}^2 x_{24}^2}$, $(1-z)(1-\bar z)=\frac{x_{14}^2 x_{23}^2}{x_{13}^2 x_{24}^2}$ are conformally-invariant cross-ratios. Conformal symmetry alone does not fix the form of the function $\GG(z,\bar z)$, which is theory-dependent. However, crossing symmetry under the exchange $x_2\leftrightarrow x_4$ implies
\begin{equation}
    \label{introab}
    \mathcal{G}(z,\bar z) = \left(\frac{z\bar z}{(1-z)(1-\bar z)}\right)^{\Delta_\phi} \mathcal{G}(1-z,1-\bar z)\,,
\end{equation}
which turns out to put surprisingly powerful restrictions on the CFT data appearing in the operator product expansion (OPE) decomposition of $\GG(z,\bar z)$. These constraints are especially powerful when combined with positivity requirements following from unitarity \cite{Poland:2018epd}.

In typical analyses of these constraints, the focus lies on restricted sets of data, and the specific form of $\GG(z,\bar z)$ is a secondary object, which is rarely accessible for generic values of $z,\bar z$. As a result, much of the detailed functional structure of the correlator remains unexplored. In this Letter, we show that neural networks (NNs) can efficiently parametrise these functions and reconstruct them from minimal input.

In what follows, we focus on the diagonal kinematics of CFT correlators where $z=\bar z$. Then, $\GG(z)=\GG(z,z)$ is a simpler, single-variable function satisfying the constraint
\begin{equation}
    \label{introac}
    \mathcal{G}(z)=\left(\frac{z}{1-z}\right)^{2\Delta_\phi}\mathcal{G}(1-z)\,.
\end{equation}
The diagonal kinematics simplify the analysis while maintaining many of the non-trivial theory-dependent features of the correlator. This has motivated in recent years considerable work on four-point correlators on the line within the conformal bootstrap approach (see e.g.~\cite{Mazac:2018mdx,Ghosh:2025sic}). In special cases, such as the $\sigma$ correlator of the 3d Ising model, it is possible to express the correlator in terms of a few Polyakov blocks that efficiently approximate it across the full kinematic domain \cite{Paulos:2020zxx}. However, constructing those explicit Polyakov blocks is technically challenging and does not generalise to generic correlators.

Along these lines, our primary focus will be the non-perturbative computation of $\GG(z)$ in the $s$-channel OPE convergence region, $z\in (0,1)$. We note, however, that our approach can be extended to the computation of the full function $\GG(z,\bar z)$ on the plane, as discussed in Section \ref{outlook} and Ref.~\cite{GKNS:1}.

\paragraph*{\bf A novel observation}
In this Letter, we report compelling evidence that a surprisingly accurate approximation of $\GG(z)$ can be reconstructed numerically on the real interval $(0,1)$ from a minimal set of data by suitably training simple feed-forward neural networks (NNs) (multi-layer-perceptron (MLP) models). These data comprise only the external scaling dimension $\Delta_\phi$, the leading behaviour of $\GG(z)$ near $z=0$, and a single value $\GG(z_0)$ at some $z_0\in (0,1)$ (chosen arbitrarily in the vicinity of 0.3). The computation is cheap and fast. The details of the approach are discussed in Section \ref{setup}. We have collected evidence by considering many examples.

\paragraph*{\bf Four-point functions}
We have studied separately contact and one-loop Witten diagrams for $\phi^4$ scalar field theory in $\text{AdS}_2$, generalised free fields (GFFs), (unitary and non-unitary) 2d minimal models, the 3d Ising model and correlators of half-BPS operators in 4d $\NN=4$ super-Yang--Mills (SYM) theory. In \cite{GKNS:3}, we also consider mixed correlators of half-BPS operators on the half-BPS Wilson line in 4d $\NN=4$ SYM theory and mixed correlators in 2d chiral algebras arising from the chiral twist of half-BPS operators in 6d (2,0) superconformal field theory.

\paragraph*{\bf Thermal two-point functions}
We find evidence that the proposed NN approach extends to other types of correlators. For instance, scalar two-point functions at finite temperature on $S^1_\beta \times \IR^{d-1}$ at zero spatial separation satisfy Eq.~\eqref{introac} as a consequence of the Kubo--Martin--Schwinger (KMS) condition \cite{Kubo1957,MartinSchwinger1959}. Applying the same approach here, we recover thermal two-point functions in several explicit cases: GFFs, generic 2d CFTs, the $\langle \sigma \sigma\rangle$ and $\langle \epsilon \epsilon\rangle$ correlators in the 3d Ising model. The latter have been studied recently in the context of the finite temperature bootstrap, e.g.~\cite{Barrat:2025wbi,Barrat:2025nvu}. We provide a novel perspective on these correlators, paving the way for new predictions and a general reorganisation of thermal bootstrap computations \cite{Iliesiu:2018fao, Petkou:2018ynm}.

\paragraph*{\bf 2d modular bootstrap}
In \cite{GKNS:2} we provide evidence for the applicability of the approach in the 2d modular bootstrap on the torus and annulus.

In all cases, we recover approximate correlators from minimal input with relative error of up to a few percent. In this Letter, we highlight some of the evidence. More examples and details appear in \cite{GKNS:1,GKNS:2,GKNS:3}.

\section{Correlation functions as MLPs}
\label{setup}

Our approach has been inspired heavily by the NN-parametrisation of thermal two-point functions in the deep finite-temperature bootstrap \cite{Niarchos:2025cdg}. Our primary interest are line-restricted correlators in the interval $(0,1)$ obeying Eq.~\eqref{introac}. The scaling dimension $\Delta_\phi$ can be positive or negative, allowing for the possibility of non-unitary CFTs.

\paragraph*{\bf Setup}
We aim to determine functions $\GG(z)$ obeying Eq.\ \eqref{introac} in conjunction with the following assumptions.

\paragraph*{\bf (a)\customlabel{para:a}{\textbf{(a)}}}
Assume $\GG(z)=L(z) + H(z)$, where $L(z)$ is a known function dominating over $H(z)$ in the limit $z\to 0$. In the OPE decomposition of $\GG(z)$, the function $L(z)$ is designed to capture a finite set of leading-order contributions from low-scaling dimension operators ($L$ stands for `low'). The function $H(z)$ is designed to capture the remainder that includes OPE contributions from arbitrarily high-energy operators ($H$ stands for `high'). $H(z)$ is the unknown target. In most examples below, we include only the universal identity contribution in $L(z)$ by setting $L(z)=1$.

\paragraph*{\bf (b)\customlabel{para:b}{\textbf{(b)}}}
Assume that the leading-order behaviour of $H(z)$ is known in the limit $z\to 0$. When this behaviour is power-like, near $z=0$ we assume that $H(z) \sim z^\delta$ with an unknown coefficient. In CFT, this is equivalent to setting a specific gap of scaling dimensions in the spectrum of operators contributing to the OPE.

\paragraph*{\bf (c)\customlabel{para:c}{\textbf{(c)}}}
Assume that the value of $H(z)$ is known at some fixed point $z_0\in (0,1)$. We call the known value $H(z_0)=H_0$ an \textit{anchor point}. From a CFT perspective, the equation $H(z_0)=H_0$ is a sum-rule for the infinite set of CFT data contributing to the OPE.

Naively, one might hope that, in this problem, partial low-energy information combined with a sum rule might guide the search towards some CFT correlator of interest. However, the above problem is severely under-determined and trivially admits an infinite number of solutions. Indeed, if $H_1(z)$ is a solution, then any function of the form $H_2(z) = H_1(z) + (1-z)^{-2\Delta_\phi} f(z)$ is also a solution as long as $f$ has the properties $f(z)=f(1-z)$, $f(z_0)=0$ and $\lim_{z\to 0}(z^{-\delta}f(z))=0$. Clearly, there is an infinite set of choices for $f(z)$. Suitable linear combinations of CFT correlators of different theories can also contribute to the indeterminacy of the problem.

On the other hand, CFT correlators are not generic solutions of the above problem. For example, four-point functions on a line are known to have an analytic continuation on the complex plane (away from two branch cuts at $(-\infty,0]\cup[1,\infty)$) and to obey appropriate boundedness conditions at infinity. Consequently, CFT correlators are typically smooth functions on the real interval $(0,1)$. Whether this implies subtle features, quantifiable with a precise mathematical criterion, is unclear.

Therefore, we should be searching for solutions of the above problem in some restricted class of functions. We can ask whether this can be achieved with a suitable parametrisation of the unknown functions $H(z)$ and whether there is a search algorithm that is implicitly or explicitly biased towards functions relevant in CFT.

\paragraph*{\bf Spectral bias in neural networks}
At this stage, it is useful to highlight some well-known facts about NNs in computer science. There are established universal approximation theorems, \cite{Cybenko1989,Hornik1989,Funahashi1989,Leshno1993}, explaining how NNs approximate generic functions and fast computer implementations optimising NNs at generic tasks. How one obtains a function of a certain type (that fits a dataset, or obeys a set of equations) is, first and foremost, encoded explicitly in the architecture of the NN and a loss function(al), whose minima are the configurations of interest. Usually, the optimisation is performed with stochastic gradient descent (SGD) based methods (e.g.\ momentum and adaptive learning methods like {\tt Adam} \cite{Kingma:2014}).

Typically, this process induces some indirect bias, which restricts the space of functions the optimiser %effectively
explores, favouring some configurations over others. Alignment of this inductive bias with the task at hand can increase the generalisation capacities of the NN.

More concretely, in gradient-based optimisation, it is known that NNs learn low-frequency components of target functions before high-frequency components, and are biased towards smooth, low-frequency, low-complexity functions, a phenomenon known as \textit{spectral bias} or \textit{frequency principle} \cite{rahaman2019spectral,Xu:2019frequency}. In the infinite width limit, the neural tangent kernel (NTK) framework \cite{Jacot:2018ntk}, quantifies exactly how this bias works and how it is affected by several factors, e.g.\ the type of activation functions. For instance, fully connected MLPs with smooth activation functions like tanh and GELU, are naturally biased towards analytic functions %that
minimising the reproducing kernel Hilbert space (RKHS) norm \cite{bietti2019inductive,gunasekar2018characterizing,arora2019fine}.

In our context, these observations raise an interesting question: could it be that a representation of line-restricted correlators by an MLP, suitably optimised on crossing symmetry and conditions \ref{para:a}-\ref{para:c}, carries the same spectral bias that underlies the illusive subclass of physical CFT correlators? The evidence we report in this paper suggests a very surprising affirmative answer.

\paragraph*{\bf A PINN formulation of 1d CFT correlators}
Motivated by this discussion, we proceed to represent the unknown part of the CFT correlator, $H(z)$, by a fully connected MLP. We have found that a light MLP with 2 or 3 layers and a relatively small width, 64 or 128, is enough for our purposes. We choose a smooth activation function (tanh or GELU), and optimise with {\tt Adam}
with a fixed-step learning schedule with decreasing learning rates from $5\times 10^{-4}$ down to $10^{-5}$--$10^{-6}$.

The loss function (typically chosen as mean square) includes two main contributions: (1) A loss enforcing the crossing equation \eqref{introac} with $\GG(z)=L(z)+H(z)$ in condition \ref{para:a}. $H(z)$ is expressed as $z^\delta {\rm NN}_{\boldsymbol{\theta}}(z)$, where ${\rm NN}_{\boldsymbol{\theta}}(z)$ is a single-variable MLP with optimisable parameters denoted collectively by a vector $\boldsymbol{\theta}$. The prefactor $z^\delta$ introduces an implicit bias towards functions satisfying \ref{para:b}. (2) A quadratic term of the form $(H(z_0)-H_0)^2$ that enforces the anchor point condition \ref{para:c}.

In this manner, we convert the search of crossing-symmetric line-restricted correlators with a fixed anchor point to a physics-informed NN (PINN). The choice of the architecture and training parameters reported here are specifically designed to encourage the phenomenon of spectral bias, which, we believe, lies at the heart of the observed alignment between the NN predictions and physical CFT correlators. We refer the reader to \cite{GKNS:1} for a more detailed discussion of this aspect.

\elsparagraph*{\bf CFT correlators from a variational principle?} 
The non-trivial element of this construction is not the  explicit (numerical) solution of crossing symmetry per se, which is ambiguous on its own, but the emergent experimental observation that the spectral bias of NN optimisation can align with subtle, potentially universal properties of CFT correlators. In all examples we have investigated, the PINN optimisation leads to surprisingly accurate approximations of physical CFT correlators, singling them out over generic crossing-symmetric functions. The numerical discrepancy with exact
results is below a few percent, and comparable with the statistical variation of our results. The small statistical variation of our runs is a combination of the stabilising nature of the anchor point and spectral bias. From a physics perspective, the outcome is very surprising and powerful. An anchor point and a gap alone allow us to characterise almost uniquely a wide range of physical correlators. This unorthodox approach bypasses many of the difficulties of traditional computational methods and suggests a unifying theme for CFTs across theories and dimensions.

The minimisation of the RKHS norm in the NTK framework implies that this unifying theme may be the existence of a {\it universal} norm that physical CFT correlators minimise. In \cite{GKNS:1}, we consider several examples of CFT correlators and crossing-symmetric deformations thereof. We study Chebyshev spectra and compare the values of several measures of smoothness---fractional Sobolev semi-norms and curvature functionals. We also perform Chebyshev--Tikhonov fits of correlation functions. This investigation confirms the special nature of physical CFT correlators among other crossing-symmetric functions and aligns with the intuition emerging from the observations based on PINN optimisation. Nevertheless, it has not allowed us yet to distil an \textit{exact} variational principle for correlators in CFT.

\section{Examples}

In this section we highlight the application of the proposed method in a few characteristic, non-trivial examples in 1, 2, 3 and 4 dimensions. All results reported below are based on the statistics collected from 100 independent runs on King's College London's CREATE~\cite{CREATE} computing cluster. In each example, we report the mean square training loss and the standard deviation of the statistics, as well as the maximum value of the mean relative error (max-RE) between the prediction and exact result over $z\in[0.01,0.9]$. We also report the corresponding maximum over $z\in[0.01,0.9]$ of the relative standard deviation of the prediction (max-Rstd) and the mean over $z$ of the statistical mean of the relative error (mean-RE). The relative error is defined as $([{\rm Prediction}] - [{\rm Exact~value}])/(1 + |[{\rm Exact~value}]|)$. The 1 at the denominator regularises the divergence at very small exact values. Plots are collected in \ref{app:plots}.

\paragraph*{\bf 1d: $\mathbf{AdS}_{\boldsymbol{2}}$ scalar}
In our first example, we consider Witten diagrams of $\phi^4$ scalar field theory in $\text{AdS}_2$. These diagrams capture holographically the perturbative expansion of correlators of a dual 1d CFT in a large central charge limit. In these examples, the leading behaviour near $z=0$ of each correction to the correlator involves logarithms (rather than pure power laws).

For the tree-level $\phi^4$ contact Witten diagram at external dimension $\Df=1$, the exact expression is
\begin{equation}
    \label{adsaa}
    \GG_{\rm contact}(z) = 2z^2\left(\frac{\log(1-z)}{z}+\frac{\log z}{1-z}\right).
\end{equation}
We set $L(z) = 2(\log z-1)\,z^2$, which captures the leading small-$z$ behaviour of \eqref{adsaa} and attempt to reconstruct the remainder $H(z)=\GG_{\rm contact}(z)-L(z)$ %, which is
parametrised as
\begin{equation}
    \label{adsab}
    H(z) = \big( z^3 \log z + z^3\log(1-z)\big)\,\text{NN}_{\boldsymbol\theta}(z)\,.
\end{equation}
With an anchor point at $z_0=0.4$ and $H(z_0)$ assumed as input from \eqref{adsaa}, we obtain the results plotted in Fig.\ \ref{fig:AdS2_contact_comparison} in \ref{app:plots}. The statistics of the NN predictions, and their comparison to the exact tree-level correlator, are
\vspace{-0.3cm}
\begin{center}
\resizebox{\columnwidth}{!}{%
\begin{tabular}{|c|c|c|c|}
\hline
Crossing Loss & max-RE & max-Rstd & mean-RE \\
\hline
$(7.19\pm 2.82)\times 10^{-8}$ & 1.3\% & 0.3\% & -0.3\% \\
\hline
\end{tabular}}
\end{center}

For the one-loop scalar bubble diagram in $\text{AdS}_2$, which is summarised in \ref{app:bubble}, the exact expression is more complicated and involves polylogarithms up to $\text{Li}_4$. Isolating the leading small-$z$ behaviour, we set
\begin{align}
    \label{adsac}
    L(z) &= \frac{27}{250} \bigg( 95 + 2\pi^4 - 80 \log z + 30 (\log z)^2
    \nonumber\\
    &\qquad\qquad+ 80 \zeta(3) - 200\zeta(3) \log z  \bigg) z^2
    \,,
\end{align}
and $H(z) = \big(z^3 \log z - z^2(\log(1-z))^2\big)\,\text{NN}_{\boldsymbol\theta}(z)$. With an anchor point at $z_0=0.4$ (informed from Eq.~\eqref{adsbubble}) we obtain the results in Fig.\ \ref{fig:AdS2_one_loop_comparison}
and the statistics
\vspace{-0.3cm}
\begin{center}
\resizebox{\columnwidth}{!}{%
\begin{tabular}{|c|c|c|c|}
\hline
Crossing Loss & max-RE & max-Rstd & mean-RE \\
\hline
$(1.19\pm 0.64)\times 10^{-6}$ & 0.6\% & 0.4\% & -0.2\% \\
\hline
\end{tabular}}
\end{center}

\paragraph*{\bf 2d: Lee--Yang minimal model}
As a 2d CFT example, we report results on the 2d Lee--Yang minimal $\MM(2,5)$, which is a non-unitary CFT. Several examples in the unitary series $\MM(p,p+1)$ can be found in \cite{GKNS:1}. We focus on the four-point correlator of the primary field $\phi_{1,2}$ with scaling dimension $\Delta_\phi=-\frac{2}{5}$, whose exact expression can be found in \ref{app:minimal}. In this example, we set $L(z)=(1-z)^{\frac45}$, corresponding to the identity-block contribution in the crossed ($t$-) channel, and parametrize the remainder $H(z)=\GG(z)-L(z)$ as $H(z) = z^{-\frac25} (1-z)^{\frac25} \, {\rm NN}_{\boldsymbol{\theta}}(z)$. With an anchor at $z_0=0.7$, we obtain the results in Fig.\ \ref{fig:lee-yang_whole_correlator_comparison}
and statistics
\vspace{-0.3cm}
\begin{center}
\resizebox{\columnwidth}{!}{%
\begin{tabular}{|c|c|c|c|}
\hline
Crossing Loss & max-RE & max-Rstd & mean-RE \\
\hline
$(2.63\pm2.35)\times 10^{-7}$ & 1.0\% & 1.3\% & $-0.2$\% \\
\hline
\end{tabular}}
\end{center}

\paragraph*{\bf 3d: Ising CFT}
High precision numerical results have been obtained for the 3d Ising CFT using the linear functional method in the conformal bootstrap \cite{Kos:2014bka,Kos:2016ysd,Simmons-Duffin:2016wlq,Chang:2024whx}. For instance, such methods have determined the values of the scaling dimensions of the $\sigma$ and $\epsilon$ operators as $\Delta_\sigma = 0.518148806(24)$, $\Delta_\epsilon = 1.41262528(29)$. Approximate expressions of the four-point correlators $\langle \sigma\sigma\sigma\sigma\rangle$ and $\langle \epsilon \epsilon\epsilon\epsilon\rangle$ can be obtained by evaluating OPE truncations using the available numerical CFT data \cite{Simmons-Duffin:2016wlq} or, independently, with the fuzzy sphere approach \cite{Zhu:2022gjc,Han:2023yyb,Hu:2023xak}. Setting $L(z)=1$ and approximating the anchor point at $z_0=0.3$ with an OPE truncation using OPE data from \cite{Simmons-Duffin:2016wlq}, we obtain the correlator $\langle \sigma\sigma\sigma\sigma\rangle$ on the line appearing in Fig.~\ref{fig:sigma_four_pt_function_anchor_0.3_comparison} and the statistics
\vspace{-0.3cm}
\begin{center}
\resizebox{\columnwidth}{!}{%
\begin{tabular}{|c|c|c|c|}
\hline
Crossing Loss & max-RE & max-Rstd & mean-RE \\
\hline
$(1.25\pm 0.15)\times 10^{-7}$ & 0.07\% & 0.02\% & $-0.02\%$ \\
\hline
\end{tabular}}
\end{center}
Here, the relative error reflects the deviation of the prediction from the OPE truncation.

\begin{figure}[ht]
	    \centering
	    \begin{tikzpicture}
	        \begin{axis}[
	            width=\columnwidth, height=5.5cm,
	            xlabel={$z$},
            ylabel={$\mathcal{G}(z)$},
            title={Ensemble Predictions vs Bootstrap},
            grid=major,
            legend pos=north west
        ]
            % Exact Solution
            \addplot [black, dashed, thick] table [x=z, y=Bootstrap, col sep=space] {plot_data/epsilon_four_pt_function_anchor_0.3_ensemble_comparison.dat};
            \addlegendentry{Bootstrap}

            % Ensemble Mean
            \addplot [blue, thick] table [x=z, y=Mean, col sep=space] {plot_data/epsilon_four_pt_function_anchor_0.3_ensemble_comparison.dat};
            \addlegendentry{Ensemble Mean}

            % Uncertainty Band
            \addplot [forget plot, name path=upper, draw=none] table [x=z, y=Mean_plus_Std, col sep=space] {plot_data/epsilon_four_pt_function_anchor_0.3_ensemble_comparison.dat};
            \addplot [forget plot, name path=lower, draw=none] table [x=z, y=Mean_minus_Std, col sep=space] {plot_data/epsilon_four_pt_function_anchor_0.3_ensemble_comparison.dat};
            \addplot [forget plot, fill=blue!30, fill opacity=0.5, draw=none] fill between [of=upper and lower];
            \addlegendimage{legend image code/.code={\fill[blue!30, draw=blue!50] (0cm,-0.1cm) rectangle (0.6cm,0.1cm);}}
            \addlegendentry{Mean $\pm$ 1 Std}
        \end{axis}
    \end{tikzpicture}
    \caption{%Plot of
    $\mathcal{G}(z)$ for %the
    $\langle \epsilon\epsilon\epsilon\epsilon\rangle$ %correlator
    in the 3d Ising model.}
    \label{fig:epsilon_four_pt_function_anchor_0.3_ensemble_comparison}
\end{figure}
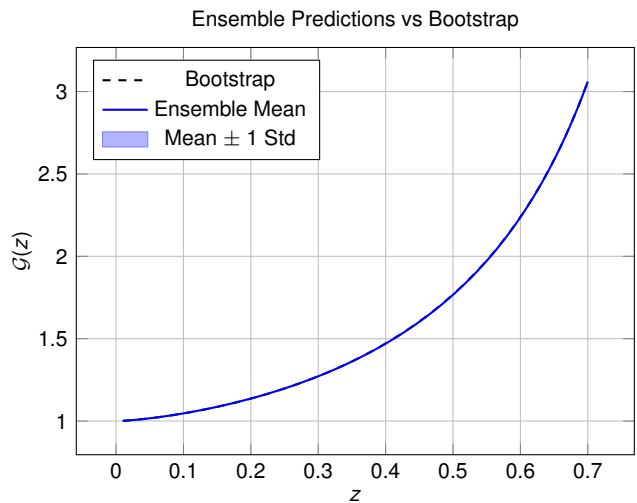

Similarly, for the $\langle \epsilon \epsilon\epsilon\epsilon\rangle$ correlator we obtain the results appearing in Fig.\ \ref{fig:epsilon_four_pt_function_anchor_0.3_ensemble_comparison}. In this plot we compare the NN prediction against an OPE truncation based again on the OPE data of \cite{Simmons-Duffin:2016wlq}. The same truncation was used to determine the anchor point at $z_0=0.3$. In this case we also evaluated the correlator at $z=0.5$ in \cite{GKNS:1} using the fuzzy sphere method \cite{Han:2023yyb}. The OPE truncation gives the correlator value $3.927$, our anchored NN $3.987 \pm 0.027$ and the fuzzy sphere $4.04$. The NN statistics %of the NN runs in this case
are
\vspace{-0.3cm}
\begin{center}
\resizebox{\columnwidth}{!}{%
\begin{tabular}{|c|c|c|c|}
\hline
Crossing Loss & max-RE & max-Rstd & mean-RE \\
\hline
$(2.42\pm 0.79)\times 10^{-6}$ & 2.4\% & 0.6\% & 0.5\% \\
\hline
\end{tabular}}
\end{center}

\paragraph*{\bf 3d: Thermal Ising CFT}

\begin{figure}[ht]
    \centering
	    \begin{tikzpicture}
	        \begin{axis}[
	            width=\columnwidth, height=5.5cm,
	            xlabel={$z$},
	            ylabel={$\mathcal{G}(z)$},
	            title={Thermal $\langle \epsilon\epsilon\rangle_\beta$ in 3d Ising},
	            grid=major,
                legend style={
                    at={(0.5,0.98)},
                    anchor=north,
                },
	            name=mainaxis,
	        ]
            \addplot [black, dashed, thick]
            table [x=z, y=Exact, col sep=space]
            {plot_data/thermal_epsilon_ensemble_comparison.dat};
            \addlegendentry{ATB}

            \addplot [blue, thick]
            table [x=z, y=Mean, col sep=space]
            {plot_data/thermal_epsilon_ensemble_comparison.dat};
            \addlegendentry{Ensemble mean}

            \addplot [forget plot, name path=upper, draw=none]
            table [x=z, y=Mean_plus_Std, col sep=space]
            {plot_data/thermal_epsilon_ensemble_comparison.dat};
            \addplot [forget plot, name path=lower, draw=none]
            table [x=z, y=Mean_minus_Std, col sep=space]
            {plot_data/thermal_epsilon_ensemble_comparison.dat};
            \addplot [forget plot, fill=blue!30, fill opacity=0.5, draw=none]
            fill between [of=upper and lower];
            \addlegendimage{legend image code/.code={\fill[blue!30, draw=blue!50] (0cm,-0.1cm) rectangle (0.6cm,0.1cm);}}
            \addlegendentry{Ensemble uncertainty band}

            % Monte Carlo uncertainty band (mean +/- quoted error)
            \addplot [forget plot, name path=mcupper, draw=none]
            table [x=z, y=MC_Mean_plus_Std, col sep=space]
            {plot_data/thermal_epsilon_mc_comparison.dat};
            \addplot [forget plot, name path=mclower, draw=none]
            table [x=z, y=MC_Mean_minus_Std, col sep=space]
            {plot_data/thermal_epsilon_mc_comparison.dat};
	            \addplot [forget plot, fill=red!20, fill opacity=0.5, draw=none]
	            fill between [of=mcupper and mclower];
			            \addlegendimage{legend image code/.code={
                            \fill[red!20, draw=red!50] (0cm,-0.1cm) rectangle (0.6cm,0.1cm);
                            \draw[red!70!black, line width=0.4pt] (0cm,0cm) -- (0.6cm,0cm);
                        }}
			            \addlegendentry{MC $\pm$ error}

                    % Interpolating curve through the MC mean points (no legend entry)
                    \addplot [forget plot, red!70!black, thin, smooth, no marks]
                    table [x=z, y=MC_Mean, col sep=space]
                    {plot_data/thermal_epsilon_mc_comparison.dat};

		            % Coordinate used for guide lines to the inset (MC mean at z=0.5)
		            \coordinate (epsMcPoint) at (axis cs:0.5,16.37863413853428);

	        \end{axis}

	        % Inset: zoom at z=0.5 showing MC band/mean, ATB, and ensemble mean.
		        \begin{axis}[
		            at={(mainaxis.center)},
		            anchor=center,
		            yshift=-4mm,
		            name=epsInset,
		            width=3.2cm,
		            height=2.5cm,
		            xmin=0.495, xmax=0.505,
		            xtick={0.5},
		            xticklabels={$0.5$},
		            % Zoom to resolve ATB vs ensemble vs MC mean at z=0.5 (MC band is clipped).
		            ymin=14, ymax=30,
	            ytick={15,20,25,30},
	            ticklabel style={font=\sansmath\sffamily\scriptsize},
	            label style={font=\sansmath\sffamily\scriptsize},
	            title style={font=\sansmath\sffamily\scriptsize},
	            % no inset title (avoid covering main plot)
	            grid=major,
	        ]
	            % MC mean +/- quoted error at z=0.5 (band is clipped by inset y-range).
	            \addplot [forget plot, name path=mcupper_inset, draw=none]
	            coordinates {(0.495,84.4967449161641) (0.505,84.4967449161641)};
	            \addplot [forget plot, name path=mclower_inset, draw=none]
	            coordinates {(0.495,-51.7394766390955) (0.505,-51.7394766390955)};
	            \addplot [forget plot, fill=red!20, fill opacity=0.5, draw=none]
	            fill between [of=mcupper_inset and mclower_inset];
	            \addplot [red!70!black]
	            coordinates {(0.495,16.3786341385343) (0.505,16.3786341385343)};

	            % ATB (Exact) at z=0.5: 28.342363...
	            \addplot [black, dashed, thick]
	            coordinates {(0.495,28.3423632811776) (0.505,28.3423632811776)};

		            % Ensemble mean at z=0.5: 21.730898...
		            \addplot [blue, thick]
		            coordinates {(0.495,21.7308983336527) (0.505,21.7308983336527)};

		            % Bottom corners of the inset plot frame (not the outer bounding box)
		            \coordinate (epsInsetBL) at (rel axis cs:0,0);
		            \coordinate (epsInsetBR) at (rel axis cs:1,0);
		        \end{axis}

		        % Guide lines (kept light so they don't obscure curves)
		        \draw[gray!80, line width=0.5pt, opacity=0.55]
		            (epsMcPoint) -- (epsInsetBL);
		        \draw[gray!80, line width=0.5pt, opacity=0.55]
		            (epsMcPoint) -- (epsInsetBR);

		    \end{tikzpicture}
	    \caption{Thermal two-point function $\GG(z)=z^{2\Delta_\epsilon}\langle\epsilon\epsilon\rangle_\beta$ in the 3d Ising CFT. %reconstructed by anchored NNs.
        We compare against the analytic thermal bootstrap (ATB) approximation and Monte Carlo (MC) data in \cite{Barrat:2025nvu}. The inset zoom at $z=0.5$ compares the MC mean and error band with the ATB value and the NN ensemble mean.}
    \label{fig:thermal_epsilon_short}
\end{figure}
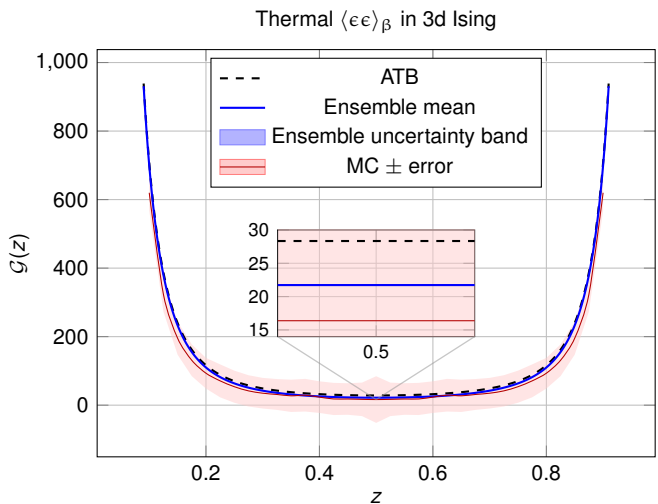

There is also evidence, \cite{GKNS:1}, that anchored NNs can similarly reconstruct thermal two-point functions of CFTs on $S^1_\beta \times \IR^{d-1}$. Again, crossing symmetry plays a central role. In particular, at zero spatial separation a suitably defined thermal two-point correlator $\GG(z)$ obeys Eq.~\eqref{introac} as a consequence of the KMS condition (precise conventions are summarised in \ref{app:thermal}). Here, as an illustration, we consider the thermal two-point functions $\GG_\sigma(z)=z^{2\Delta_\sigma} \langle \sigma \sigma\rangle_\beta$ and $\GG_\epsilon(z)=z^{2\Delta_\epsilon} \langle \epsilon \epsilon\rangle_\beta$ in the 3d Ising CFT at $\beta=1$. $\Delta_\sigma, \Delta_\epsilon$ are the values quoted above. In this case, the exact two-point functions are unknown, and an anchor point is not readily available. We approximate it at $z_0=0.05$ using the OPE contribution of the leading non-trivial thermal OPE coefficient $a_\epsilon$ of the $\epsilon$ operator. The latter has been estimated at 0.75(15) for $\langle \sigma\sigma\rangle_\beta$ and 1.09(22) for $\langle \epsilon\epsilon\rangle_\beta$ in \cite{Barrat:2025wbi}. In Fig.~\ref{fig:thermal_epsilon_short} we present results for the more demanding $\langle \epsilon\epsilon\rangle_\beta$ correlator comparing with the analytic approximation and independent Monte Carlo results in \cite{Barrat:2025wbi,Barrat:2025nvu}. In Fig.~\ref{fig:thermal_sigma_short_app}
we summarise the corresponding results for $\langle \sigma\sigma\rangle_\beta$.

\paragraph*{\bf 4d: \boldmath{$\NN\!=\!4$} SYM theory}

$\NN=4$ SYM theory is a much-studied four-dimensional supersymmetric gauge theory. The energy-momentum tensor belongs in a supersymmetric multiplet with superconformal primary scalar operators in the $\boldsymbol{20}'$ representation of the $\mathfrak{so}(6)_R$ R-symmetry algebra of the theory. Superconformal Ward identities can be used to recast the four-point function of these scalar operators in terms of a single function of the cross-ratios $\HH(z,\bar z)$. On a line, the function $\HH(z)=\HH(z,z)$ obeys the crossing equation
\begin{equation}
    \label{symaa}
    (1-z)^2 \HH(z)-z^2 \HH(1-z)+\frac{2z-1}{c}+z^4-(1-z)^4=0\,,
\end{equation}
where $c$ is the central charge of the theory. In the large-$c$ limit, $\HH(z)$ has a $1/c$-expansion of the form $\HH^{(0)}(z) + \frac{1}{c} \HH^{(1)}(z) +\OO(c^{-2})$ with $\HH^{(0)}(z)=1+(1-z)^{-4}$ and $\HH^{(1)}(z)=(1-z)^{-2}+z^4 \bar D_{2422}(z)$. The function $\bar D_{2422}(z)$ is given in Eq.~\eqref{n4symac}. Using anchored NNs at $z_0=0.3$ with $L(z)=1$ and $\HH^{(1)}(z)=z (1-z)^{-2}{\rm NN}_{\boldsymbol{\theta}}(z)$, we obtain the results in Fig.~\ref{fig:AdS2_contact_summary} of \ref{app:plots} and the statistics
\vspace{-0.3cm}
\begin{center}
\resizebox{\columnwidth}{!}{%
\begin{tabular}{|c|c|c|c|}
\hline
Crossing Loss & max-RE & max-Rstd & mean-RE \\
\hline
$(8.06\pm 7.02)\times 10^{-8}$ & 0.8\% & 1.7\% & 0.06\% \\
\hline
\end{tabular}}
\end{center}

\section{Outlook}
\label{outlook}

In this Letter we reported intriguing observations relating conformal correlators on a line to the optimisation of neural networks. A more detailed exposition with additional examples appears in \cite{GKNS:1,GKNS:2,GKNS:3}. Our main goal has been the concrete demonstration of the constraining power of a 'crossing-anchor-smoothness' combination on physical CFT correlators.

In all reported examples, we assumed input corresponding to physical correlators and checked how the predicted output compares to known information. Our method is constructive and will always produce a prediction even if the input is unphysical. In examples of fake anchor input with extremely high values, which are expected to lead to unphysical correlators, we have observed that the tension between smoothness and crossing-symmetry leads to outputs with high training loss. However, without any further information, it is impossible at the current stage of development to reject unphysical input or predictions, or to distinguish different types of physical correlators satisfying the same minimal set of assumptions.   

Our observations suggest the existence of a hitherto unknown \textit{universal} principle underlying CFT correlators involving the minimisation of some (potentially non-local) functional. In \cite{GKNS:1}, we explore this aspect further, testing the potential relevance of a curvature functional and the fractional Sobolev semi-norm.  We also introduce alternative non-neural parametrisations of the correlators and Chebyshev--Tikhonov optimisation schemes. Exploring further the interplay between low-lying CFT data, anchor points and the smoothness of the correlators is particularly interesting and may lead eventually to methodologies that could even bypass the need for an independent anchor-point input.

In this Letter, we emphasised the computation of conformal correlators on a line. In \cite{GKNS:1}, we present a related approach, where PINN optimisation based on crossing symmetry on concentric circles around $z=\frac{1}{2}$ allows us to reconstruct (without any further input) the full four-point functions $\GG(z,\bar z)$ in several examples. This provides preliminary evidence that anchored NNs are not limited to line-restricted correlators, but can also be used to recover four-point functions (as well as thermal two-point functions) beyond the diagonal kinematics.

We conclude by noting that the reported observations provide a new, unexpected bridge between physics and computer science. NNs have been employed in the past to reformulate problems in QFT and quantum many-body physics (NN QFTs \cite{Halverson:2020trp,Demirtas:2023fir,Halverson:2024axc,Ferko:2026axm} are a notable recent example). Our observations are qualitatively different, not only because they offer a novel flexible computational alternative in QFT (beyond the standard Lagrangian framework), but also because they employ a subtle feature of NNs to uncover a potentially deep universal aspect of CFT, paving the path towards new discoveries in directions orthogonal to the state-of-the-art. It would be very interesting to establish this methodology further as a concrete part of modern CFT and as a viable complement to existing methods for the study of QFT.

\section*{Acknowledgments}

We would like to thank A.~Kaviraj, S.~Pal, C.~Papageorgakis, A.~Stratoudakis, J.~ Thaler, and M.~Woolley for useful discussions, and J.~Barrat, D.~N.~Bozkurt, E.~Marchetto, A.~Miscioscia and E.~Pomoni for sharing their data on 3d Ising thermal correlators in Ref.\ \cite{Barrat:2025nvu}. Research presented in this work was initiated with and supported by an ``International Exchanges 2024 Global Round 1'' grant from the Royal Society (IES\textbackslash{}R1\textbackslash{}241082). Numerical computations in this work have been largely performed on King's College London's CREATE~\cite{CREATE} computing cluster. Claude Code was used in this work for code development. KG is supported by the Royal Society under grant RF\textbackslash{}ERE\textbackslash{}231142. SK is supported by the UK's Engineering and Physical Sciences Research Council under grant EP/Z535035/1, through an EPSRC Doctoral Landscape Award. AS is supported by the Royal Society under grant URF\textbackslash{}R1\textbackslash211417 and by STFC under grant ST/X000753/1.

\appendix

\section{Exact correlators}
\label{app:exact}

\subsection{Bubble diagram of scalar \texorpdfstring{$\phi^4$}{phi\^4} in \texorpdfstring{$\text{AdS}_2$}{AdS\_2}}
\label{app:bubble}

The exact one-loop Witten diagram in scalar $\phi^4$ theory in $\text{AdS}_2$ for $\Delta_{\phi}=1$ is given by \cite{Mazac:2018ycv}
\begin{align}
    \label{adsbubble}
    &\GG_{\rm bubble}(z) = \frac{9}{250\,(z-1)^{2}} \Bigg[
    \pi^{4} z^{2}\bigl(6 - (z-2)z\bigr)\nonumber\\
    &
    + 15\Bigl(
    4 z^{3}\operatorname{arctanh}(1-2z)
    + (2z-1)\log^{4}(1-z)
    \nonumber\\
    &
    + 2 z^{2}\log z
    + 4(1-2z)\log^{3}(1-z)\log z
    \nonumber\\
    &
    + 2\log(1-z)\bigl(z-2z^{2}
    + \pi^{2}(z-1)^{2}(2+z^{2})\log z\bigr)
    \nonumber\\
    &
    + 2\log^{2}(1-z)\bigl(\pi^{2}(2z-1)
    - 3(z-1)^{2}(1+z^{2})\log^{2} z\bigr)
    \Bigr)
    \nonumber\\
    &- 180 (z-1)^{2}\Bigl((z^{2}-1)\log(1-z)
    \nonumber\\
    &\hspace{3cm}+ (2+z^{2})\log z\Bigr)\operatorname{Li}_{3}(1-z)
    \nonumber\\
    &- 180 z^{2}\Bigl((3+(z-2)z)\log(1-z)
    + (z-2)z\log z\Bigr)\operatorname{Li}_{3}(z)
    \nonumber\\
    &+ 360 (z-2) z^{3}\operatorname{Li}_{4}(1-z)
    + 360 (z-1)^{3}(1+z)\operatorname{Li}_{4}(z)
    \nonumber\\
    &
    + 360(2z-1)\operatorname{Li}_{4}\!\left(\frac{z}{z-1}\right)
    \nonumber\\
    &+ 60\Bigl(-8 z^{3}\operatorname{arctanh}(1-2z)
    \nonumber\\
    &\hspace{2cm}+ (-3+z(2+11z))\log(1-z)
    \nonumber\\
    &\hspace{2cm} - (-6+z(12+z))\log z\Bigr)\zeta(3)
    \Bigg].
\end{align}

\subsection{2d minimal model correlators on a line}
\label{app:minimal}

The reduced four-point function of the primary field $\phi_{1,2}$ in the 2d minimal model $\MM(p,p')$, with scaling dimension $\Delta_\phi = \frac{3}{2}\frac{p}{p'}-1$, is \cite{Liendo:2012hy}
\begin{equation}
    \label{minaa}
    \GG(z) = (\GG_1(z))^2+ N(\Df)\, (\GG_2(z))^2
    \,,
\end{equation}
with
\begin{align}
    \label{minab}
    \GG_1(z) =& (1-z)^{-\Df}
    \nonumber\\
    &\times \,{}_2F_1\big(\tfrac13(1-2\Df),\,-2\Df;\,\tfrac23(1-2\Df);\,z\big)\,,
    \nonumber\\
    \GG_2(z) =& (1-z)^{\frac{1+\Df}{3}}\,z^{\frac{1+4\Df}{3}}\,
    \nonumber\\&\times {}_2F_1\big(\tfrac23(1+\Df),\,1+2\Df;\,\tfrac43(1+\Df);\,z\big)\,,\nonumber\\
    N(\Df) =& 2^{1-\frac{8(\Df+1)}{3}}\,\Gamma\bigl(\tfrac{2}{3}-\tfrac{4\Df}{3}\bigr)^2\,\Gamma(2\Df+1)^2\,
    \nonumber\\
    &\times \bigg(\sin\tfrac{\pi(16\Df+1)}{6}-\cos\tfrac{\pi(4\Df+1)}{3}\bigg)
    \nonumber\\
    &\times \bigg(\pi\,\Gamma\left(\tfrac{2\Df}{3}+\tfrac{7}{6}\right)^2\bigg)^{-1}
    \,.
\end{align}

\subsection{Thermal two-point functions}
\label{app:thermal}

For a finite-temperature $d$-dimensional CFT on $S^1_\beta\times \IR^{d-1}$ the Euclidean time $\tau\in[0,\beta)$ is compactified. Denoting the spacetime coordinates as $(\tau, \vec x)$, the thermal two-point function of a scalar primary operator depends only on $\tau$ and $|x|=\sqrt{\tau^2+\vec x^2}$:
\begin{equation}
    \label{thermalaa}
    g(\tau,|x|)=\langle \phi(\tau, \vec x) \phi(0,0)\rangle_\beta\,.
\end{equation}
Using the $SO(d-1)$ rotational symmetry of the CFT on $S^1_\beta\times \IR^{d-1}$, we can set $\vec x=(\rho,0,\ldots,0)$ and define $z=\tau+i\rho$, $\bar z=\tau-i\rho$. At $\beta=1$, the KMS condition (plus parity) implies the crossing equation $g(1-z,1-\bar z)=g(z,\bar z)$. In the main text, we consider scalar two-point functions at zero spatial separation $\rho=0$ (which implies $z=\bar z$) and denote $\GG(z)=z^{2\Delta_\phi} g(z,z)$. As always, $\Delta_\phi$ is the scaling dimension of the operator $\phi$.

\subsection{Correlators of half-BPS operators in 4d \texorpdfstring{$\NN=4$}{N=4} SYM}
\label{n4sym}

In the main text we considered four-point correlation functions of the half-BPS superconformal primaries in the ${\boldsymbol{20}'}$. The superconformal Ward identities can be used to express the four-point correlator in terms of a single function $\HH(z,\bar z)$ of the conformal cross-ratios. In a large-$c$ expansion, $\HH(z)=\HH^{(0)}(z) + \frac{1}{c}\HH^{(1)}(z)+\OO(c^{-2})$ with the first non-trivial correction $\HH^{(1)}(z)$ obeying the crossing equation
\begin{equation}
    \label{n4symaa}
    (1-z)^4 \HH^{(1)}(z)-z^4 \HH^1(1-z)-(1-2z)=0\,.
\end{equation}
The exact solution is \cite{Dolan:2006ec}
\begin{equation}
    \label{n4symab}
    \HH^{(1)}(z) = (1-z)^{-2}+z^4 \bar D_{2422}(z)\,,
\end{equation}
with
\begin{align}
    \label{n4symac}
    &\bar{D}_{2422}(z)= \bigg[4 (z (2 z-7)+7) z^5 \tanh ^{-1}(1-2 z)
    \nonumber\\
    &+4 (z-1) ((z-1) z+1)^2 z
    -2 (7 (z-1) z+2) \log (1-z)\bigg]
    \nonumber\\
    &\times \bigg[35 (z-1)^5 z^5\bigg]^{-1}
    .
\end{align}

\section{Prediction plots}
\label{app:plots}

In this appendix we collect plots in the five highlighted examples of the main text. We plot the anchored neural network predictions (blue solid lines) with the statistical standard deviation over 100 cluster runs (blue band, barely visible) against the exact analytic curve (dashed line).

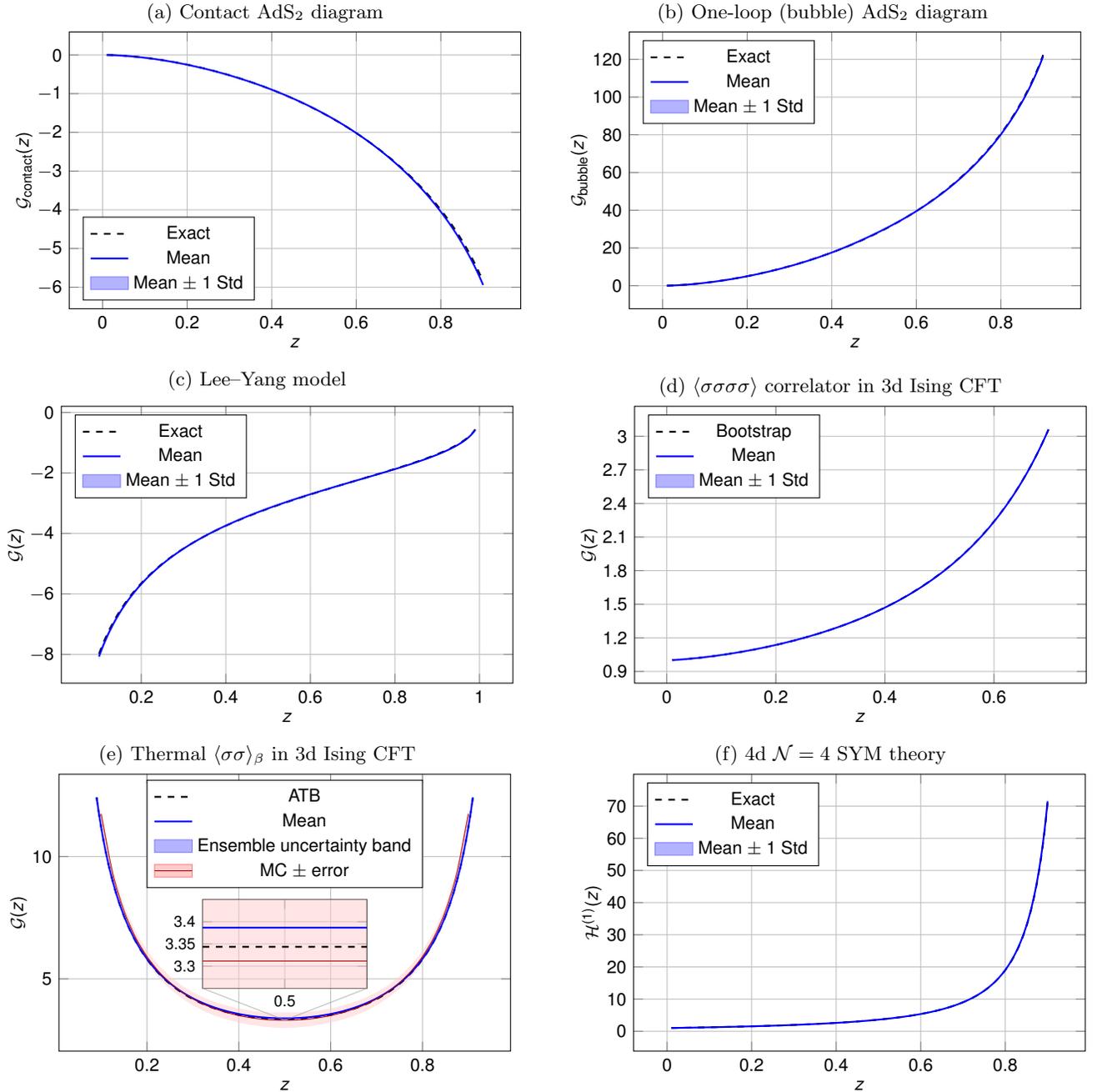
\begin{figure*}[t]
    \centering
    % Subfigure 1: Comparison
    \begin{subfigure}[b]{0.49\textwidth}
        \caption{Contact $\text{AdS}_2$ diagram}
        \label{fig:AdS2_contact_comparison}
        \centering
        \begin{tikzpicture}
            \begin{axis}[
                width=\linewidth, height=6cm,
                xlabel={$z$},
                ylabel={$\mathcal{G}_{\text{contact}}(z)$},
                grid=major,
                ytick distance=1,
                legend pos=south west,
            ]
                % Exact Solution
                \addplot [black, dashed, thick] table [x=z, y=Exact, col sep=space] {plot_data/AdS2_contact_ensemble_comparison.dat};
                \addlegendentry{Exact}

                % Ensemble Mean
                \addplot [blue, thick] table [x=z, y=Mean, col sep=space] {plot_data/AdS2_contact_ensemble_comparison.dat};
                \addlegendentry{Mean}

                % Uncertainty Band
                \addplot [forget plot, name path=upper, draw=none] table [x=z, y=Mean_plus_Std, col sep=space] {plot_data/AdS2_contact_ensemble_comparison.dat};
                \addplot [forget plot, name path=lower, draw=none] table [x=z, y=Mean_minus_Std, col sep=space] {plot_data/AdS2_contact_ensemble_comparison.dat};
                \addplot [forget plot, fill=blue!30, fill opacity=0.5, draw=none] fill between [of=upper and lower];
                \addlegendimage{legend image code/.code={\fill[blue!30, draw=blue!50] (0cm,-0.1cm) rectangle (0.6cm,0.1cm);}}
                \addlegendentry{Mean $\pm$ 1 Std}
            \end{axis}
        \end{tikzpicture}
    \end{subfigure}
    %
    %\hfill
    %
    \begin{subfigure}[b]{0.49\textwidth}
        \centering
        \caption{One-loop (bubble) $\text{AdS}_2$ diagram}
        \label{fig:AdS2_one_loop_comparison}
        \begin{tikzpicture}
            \begin{axis}[
                width=\linewidth, height=6cm,
                xlabel={$z$},
                ylabel={$\mathcal{G}_{\text{bubble}}(z)$},
                ytick distance=20,
                grid=major,
                legend pos=north west,
            ]
                % Exact Solution
                \addplot [black, dashed, thick] table [x=z, y=Exact, col sep=space] {plot_data/AdS2_one_loop_ensemble_comparison.dat};
                \addlegendentry{Exact}

                % Ensemble Mean
                \addplot [blue, thick] table [x=z, y=Mean, col sep=space] {plot_data/AdS2_one_loop_ensemble_comparison.dat};
                \addlegendentry{Mean}

                % Uncertainty Band
                \addplot [forget plot, name path=upper, draw=none] table [x=z, y=Mean_plus_Std, col sep=space] {plot_data/AdS2_one_loop_ensemble_comparison.dat};
                \addplot [forget plot, name path=lower, draw=none] table [x=z, y=Mean_minus_Std, col sep=space] {plot_data/AdS2_one_loop_ensemble_comparison.dat};
                \addplot [forget plot, fill=blue!30, fill opacity=0.5, draw=none] fill between [of=upper and lower];
                \addlegendimage{legend image code/.code={\fill[blue!30, draw=blue!50] (0cm,-0.1cm) rectangle (0.6cm,0.1cm);}}
                \addlegendentry{Mean $\pm$ 1 Std}
            \end{axis}
        \end{tikzpicture}
    \end{subfigure}

\begin{subfigure}[b]{0.49\textwidth}
        \centering
        \caption{Lee--Yang model}
        \label{fig:lee-yang_whole_correlator_comparison}
        \begin{tikzpicture}
            \begin{axis}[
                width=\linewidth, height=6cm,
                xlabel={$z$},
                ylabel={$\mathcal{G}(z)$},
                grid=major,
                ytick distance=2,
                legend pos=north west,
            ]
                % Exact Solution
                \addplot [black, dashed, thick] table [x=z, y=Exact, col sep=space] {plot_data/lee-yang_whole_correlator_ensemble_comparison.dat};
                \addlegendentry{Exact}

                % Ensemble Mean
                \addplot [blue, thick] table [x=z, y=Mean, col sep=space] {plot_data/lee-yang_whole_correlator_ensemble_comparison.dat};
                \addlegendentry{Mean}

                % Uncertainty Band
                \addplot [forget plot, name path=upper, draw=none] table [x=z, y=Mean_plus_Std, col sep=space] {plot_data/lee-yang_whole_correlator_ensemble_comparison.dat};
                \addplot [forget plot, name path=lower, draw=none] table [x=z, y=Mean_minus_Std, col sep=space] {plot_data/lee-yang_whole_correlator_ensemble_comparison.dat};
                \addplot [forget plot, fill=blue!30, fill opacity=0.5, draw=none] fill between [of=upper and lower];
                \addlegendimage{legend image code/.code={\fill[blue!30, draw=blue!50] (0cm,-0.1cm) rectangle (0.6cm,0.1cm);}}
                \addlegendentry{Mean $\pm$ 1 Std}
            \end{axis}
        \end{tikzpicture}
    \end{subfigure}
    \hfill
    \begin{subfigure}[b]{0.49\textwidth}
        \centering
        \caption{$\langle \sigma \sigma \sigma \sigma \rangle$ correlator in 3d Ising CFT}
        \label{fig:sigma_four_pt_function_anchor_0.3_comparison}
        \begin{tikzpicture}
            \begin{axis}[
                width=\linewidth, height=6cm,
                xlabel={$z$},
                ylabel={$\mathcal{G}(z)$},
                grid=major,
                ytick distance=0.3,
                legend pos=north west,
            ]
                % Exact Solution
                \addplot [black, dashed, thick] table [x=z, y=Bootstrap, col sep=space] {plot_data/sigma_four_pt_function_anchor_0.3_ensemble_comparison.dat};
                \addlegendentry{Bootstrap}

                % Ensemble Mean
                \addplot [blue, thick] table [x=z, y=Mean, col sep=space] {plot_data/sigma_four_pt_function_anchor_0.3_ensemble_comparison.dat};
                \addlegendentry{Mean}

                % Uncertainty Band
                \addplot [forget plot, name path=upper, draw=none] table [x=z, y=Mean_plus_Std, col sep=space] {plot_data/sigma_four_pt_function_anchor_0.3_ensemble_comparison.dat};
                \addplot [forget plot, name path=lower, draw=none] table [x=z, y=Mean_minus_Std, col sep=space] {plot_data/sigma_four_pt_function_anchor_0.3_ensemble_comparison.dat};
                \addplot [forget plot, fill=blue!30, fill opacity=0.5, draw=none] fill between [of=upper and lower];
                \addlegendimage{legend image code/.code={\fill[blue!30, draw=blue!50] (0cm,-0.1cm) rectangle (0.6cm,0.1cm);}}
                \addlegendentry{Mean $\pm$ 1 Std}
            \end{axis}
        \end{tikzpicture}
    \end{subfigure}

	    \begin{subfigure}[b]{0.49\textwidth}
	        \centering
	        \caption{Thermal $\langle \sigma\sigma\rangle_\beta$ in 3d Ising CFT}
		        \label{fig:thermal_sigma_short_app}
			        \begin{tikzpicture}
			            \begin{axis}[
			                width=\linewidth, height=6cm,
			                xlabel={$z$},
		                ylabel={$\mathcal{G}(z)$},
	                grid=major,
	                legend style={
                    at={(0.5,0.98)},
                    anchor=north,
                    },
	                name=mainaxisThermApp,
	            ]
                \addplot [black, dashed, thick]
                table [x=z, y=Exact, col sep=space]
	                {plot_data/thermal_sigma_ensemble_comparison.dat};
	                \addlegendentry{ATB}

	                \addplot [blue, thick]
	                table [x=z, y=Mean, col sep=space]
	                {plot_data/thermal_sigma_ensemble_comparison.dat};
	                \addlegendentry{Mean}

	                \addplot [forget plot, name path=upper, draw=none]
	                table [x=z, y=Mean_plus_Std, col sep=space]
	                {plot_data/thermal_sigma_ensemble_comparison.dat};
	                \addplot [forget plot, name path=lower, draw=none]
	                table [x=z, y=Mean_minus_Std, col sep=space]
	                {plot_data/thermal_sigma_ensemble_comparison.dat};
	                \addplot [forget plot, fill=blue!30, fill opacity=0.5, draw=none]
	                fill between [of=upper and lower];
	                \addlegendimage{legend image code/.code={\fill[blue!30, draw=blue!50] (0cm,-0.1cm) rectangle (0.6cm,0.1cm);}}
	                \addlegendentry{Ensemble uncertainty band}

                % Monte Carlo uncertainty band (mean +/- quoted error)
		                \addplot [forget plot, name path=mcupper, draw=none]
		                table [x=z, y=MC_Mean_plus_Std, col sep=space]
		                {plot_data/thermal_sigma_mc_comparison.dat};
	                \addplot [forget plot, name path=mclower, draw=none]
	                table [x=z, y=MC_Mean_minus_Std, col sep=space]
	                {plot_data/thermal_sigma_mc_comparison.dat};
			                \addplot [forget plot, fill=red!20, fill opacity=0.5, draw=none]
			                fill between [of=mcupper and mclower];
				                \addlegendimage{legend image code/.code={
                                    \fill[red!20, draw=red!50] (0cm,-0.1cm) rectangle (0.6cm,0.1cm);
                                    \draw[red!70!black, line width=0.4pt] (0cm,0cm) -- (0.6cm,0cm);
                                }}
				                \addlegendentry{MC $\pm$ error}

                                % Interpolating curve through the MC mean points (no legend entry)
                                \addplot [forget plot, red!70!black, thin, smooth, no marks]
                                table [x=z, y=MC_Mean, col sep=space]
                                {plot_data/thermal_sigma_mc_comparison.dat};

				                % Coordinate used for guide lines to the inset (MC mean at z=0.5)
				                \coordinate (sigMcPointApp) at (axis cs:0.5,3.311604420587241);

			            \end{axis}

	            % Inset: zoom at z=0.5 showing MC band/mean, ATB, and ensemble mean.
			            \begin{axis}[
			                at={(mainaxisThermApp.center)},
			                anchor=center,
				                yshift=-5mm,
				                name=sigInsetApp,
				                width=4.2cm,
				                height=3.0cm,
				                xmin=0.495, xmax=0.505,
			                xtick={0.5},
			                xticklabels={$0.5$},
				                % Zoom to resolve ATB vs ensemble vs MC mean at z=0.5 (MC band is clipped).
				                ymin=3.25, ymax=3.45,
			                ytick={3.30,3.35,3.40},
			                ticklabel style={font=\sansmath\sffamily\scriptsize},
		                label style={font=\sansmath\sffamily\scriptsize},
		                title style={font=\sansmath\sffamily\scriptsize},
		                % no inset title (avoid covering main plot)
		                grid=major,
		            ]
		                \addplot [forget plot, name path=mcupper_inset, draw=none]
		                coordinates {(0.495,3.630055130379896) (0.505,3.630055130379896)};
		                \addplot [forget plot, name path=mclower_inset, draw=none]
		                coordinates {(0.495,2.993153710794585) (0.505,2.993153710794585)};
		                \addplot [forget plot, fill=red!20, fill opacity=0.5, draw=none]
		                fill between [of=mcupper_inset and mclower_inset];
		                \addplot [red!70!black]
			                coordinates {(0.495,3.311604420587241) (0.505,3.311604420587241)};

		                \addplot [black, dashed, thick]
			                coordinates {(0.495,3.34352831402591) (0.505,3.34352831402591)};

			                \addplot [blue, thick]
			                coordinates {(0.495,3.386629959572441) (0.505,3.386629959572441)};

			                % Bottom corners of the inset plot frame (not the outer bounding box)
			                \coordinate (sigInsetAppBL) at (rel axis cs:0,0);
			                \coordinate (sigInsetAppBR) at (rel axis cs:1,0);
			            \end{axis}

			            % Guide lines (kept light so they don't obscure curves)
			            \draw[gray!80, line width=0.5pt, opacity=0.55]
			                (sigMcPointApp) -- (sigInsetAppBL);
			            \draw[gray!80, line width=0.5pt, opacity=0.55]
			                (sigMcPointApp) -- (sigInsetAppBR);

		        \end{tikzpicture}
	    \end{subfigure}
    \hfill
    \begin{subfigure}[b]{0.49\textwidth}
        \centering
        \caption{4d $\NN=4$ SYM theory}
        \label{fig:o20p_comparison}
        \begin{tikzpicture}
            \begin{axis}[
                width=\linewidth, height=6cm,
                xlabel={$z$},
                ylabel={$\mathcal{H}^{(1)}(z)$},
                grid=major,
                ytick distance=10,
                legend pos=north west,
            ]
                % Exact Solution
                \addplot [black, dashed, thick] table [x=z, y=Exact, col sep=space] {plot_data/o20p_ensemble_comparison.dat};
                \addlegendentry{Exact}

                % Ensemble Mean
                \addplot [blue, thick] table [x=z, y=Mean, col sep=space] {plot_data/o20p_ensemble_comparison.dat};
                \addlegendentry{Mean}

                % Uncertainty Band
                \addplot [forget plot, name path=upper, draw=none] table [x=z, y=Mean_plus_Std, col sep=space] {plot_data/o20p_ensemble_comparison.dat};
                \addplot [forget plot, name path=lower, draw=none] table [x=z, y=Mean_minus_Std, col sep=space] {plot_data/o20p_ensemble_comparison.dat};
                \addplot [forget plot, fill=blue!30, fill opacity=0.5, draw=none] fill between [of=upper and lower];
                \addlegendimage{legend image code/.code={\fill[blue!30, draw=blue!50] (0cm,-0.1cm) rectangle (0.6cm,0.1cm);}}
                \addlegendentry{Mean $\pm$ 1 Std}
            \end{axis}
        \end{tikzpicture}
    \end{subfigure}

    \caption{Collected plots of anchored neural network predictions against analytic results.}
    \label{fig:AdS2_contact_summary}
\end{figure*}

\clearpage
\bibliographystyle{elsarticle-num}
\bibliography{ancb_plb}

\begin{thebibliography}{10}
\expandafter\ifx\csname url\endcsname\relax
  \def\url#1{\texttt{#1}}\fi
\expandafter\ifx\csname urlprefix\endcsname\relax\def\urlprefix{URL }\fi
\expandafter\ifx\csname href\endcsname\relax
  \def\href#1#2{#2} \def\path#1{#1}\fi

\bibitem{Poland:2018epd}
D.~Poland, S.~Rychkov, A.~Vichi, {The Conformal Bootstrap: Theory, Numerical Techniques, and Applications}, Rev. Mod. Phys. 91 (2019) 015002.
\newblock \href {http://arxiv.org/abs/1805.04405} {\path{arXiv:1805.04405}}, \href {https://doi.org/10.1103/RevModPhys.91.015002} {\path{doi:10.1103/RevModPhys.91.015002}}.

\bibitem{Mazac:2018mdx}
D.~Mazac, M.~F. Paulos, {The analytic functional bootstrap. Part I: 1D CFTs and 2D S-matrices}, JHEP 02 (2019) 162.
\newblock \href {http://arxiv.org/abs/1803.10233} {\path{arXiv:1803.10233}}, \href {https://doi.org/10.1007/JHEP02(2019)162} {\path{doi:10.1007/JHEP02(2019)162}}.

\bibitem{Ghosh:2025sic}
K.~Ghosh, M.~F. Paulos, N.~Suchel, {Solving 1D crossing and QFT$_2$/CFT$_1$} (3 2025).
\newblock \href {http://arxiv.org/abs/2503.22798} {\path{arXiv:2503.22798}}.

\bibitem{Paulos:2020zxx}
M.~F. Paulos, {Dispersion relations and exact bounds on CFT correlators}, JHEP 08 (2021) 166.
\newblock \href {http://arxiv.org/abs/2012.10454} {\path{arXiv:2012.10454}}, \href {https://doi.org/10.1007/JHEP08(2021)166} {\path{doi:10.1007/JHEP08(2021)166}}.

\bibitem{GKNS:1}
K.~Ghosh, S.~Kumar, V.~Niarchos, A.~Stergiou, {Neural Spectral Bias and Conformal Correlators I: Introduction and Applications} (4 2026).
\newblock \href {http://arxiv.org/abs/2604.18686} {\path{arXiv:2604.18686}}.

\bibitem{GKNS:3}
K.~Ghosh, S.~Kumar, V.~Niarchos, A.~Stergiou, {Neural Spectral Bias and Conformal Correlators III: Chiral Algebra Twists and Bootstrability}In preparation (2026).

\bibitem{Kubo1957}
R.~Kubo, Statistical-mechanical theory of irreversible processes. i. general theory and simple applications to magnetic and conduction problems, Journal of the Physical Society of Japan 12~(6) (1957) 570--586.
\newblock \href {https://doi.org/10.1143/JPSJ.12.570} {\path{doi:10.1143/JPSJ.12.570}}.

\bibitem{MartinSchwinger1959}
P.~C. Martin, J.~Schwinger, Theory of many-particle systems. i, Physical Review 115~(6) (1959) 1342--1373.
\newblock \href {https://doi.org/10.1103/PhysRev.115.1342} {\path{doi:10.1103/PhysRev.115.1342}}.

\bibitem{Barrat:2025wbi}
J.~Barrat, E.~Marchetto, A.~Miscioscia, E.~Pomoni, {Thermal Bootstrap for the Critical O(N) Model}, Phys. Rev. Lett. 134~(21) (2025) 211604.
\newblock \href {http://arxiv.org/abs/2411.00978} {\path{arXiv:2411.00978}}, \href {https://doi.org/10.1103/PhysRevLett.134.211604} {\path{doi:10.1103/PhysRevLett.134.211604}}.

\bibitem{Barrat:2025nvu}
J.~Barrat, D.~N. Bozkurt, E.~Marchetto, A.~Miscioscia, E.~Pomoni, {The analytic bootstrap at finite temperature} (6 2025).
\newblock \href {http://arxiv.org/abs/2506.06422} {\path{arXiv:2506.06422}}.

\bibitem{Iliesiu:2018fao}
L.~Iliesiu, M.~Kologlu, R.~Mahajan, E.~Perlmutter, D.~Simmons-Duffin, {The Conformal Bootstrap at Finite Temperature}, JHEP 10 (2018) 070.
\newblock \href {http://arxiv.org/abs/1802.10266} {\path{arXiv:1802.10266}}, \href {https://doi.org/10.1007/JHEP10(2018)070} {\path{doi:10.1007/JHEP10(2018)070}}.

\bibitem{Petkou:2018ynm}
A.~C. Petkou, A.~Stergiou, {Dynamics of Finite-Temperature Conformal Field Theories from Operator Product Expansion Inversion Formulas}, Phys. Rev. Lett. 121~(7) (2018) 071602.
\newblock \href {http://arxiv.org/abs/1806.02340} {\path{arXiv:1806.02340}}, \href {https://doi.org/10.1103/PhysRevLett.121.071602} {\path{doi:10.1103/PhysRevLett.121.071602}}.

\bibitem{GKNS:2}
K.~Ghosh, S.~Kumar, V.~Niarchos, A.~Stergiou, {Neural Spectral Bias and Conformal Correlators II: Modular and Annulus Bootstrap}In preparation (2026).

\bibitem{Niarchos:2025cdg}
V.~Niarchos, C.~Papageorgakis, A.~Stratoudakis, M.~Woolley, {Deep finite temperature bootstrap}, Phys. Rev. D 112~(12) (2025) 126012.
\newblock \href {http://arxiv.org/abs/2508.08560} {\path{arXiv:2508.08560}}, \href {https://doi.org/10.1103/qrjx-w4md} {\path{doi:10.1103/qrjx-w4md}}.

\bibitem{Cybenko1989}
G.~Cybenko, Approximation by superpositions of a sigmoidal function, Mathematics of Control, Signals, and Systems 2~(4) (1989) 303--314.
\newblock \href {https://doi.org/10.1007/BF02551274} {\path{doi:10.1007/BF02551274}}.

\bibitem{Hornik1989}
K.~Hornik, M.~Stinchcombe, H.~White, Multilayer feedforward networks are universal approximators, Neural Networks 2~(5) (1989) 359--366.
\newblock \href {https://doi.org/10.1016/0893-6080(89)90020-8} {\path{doi:10.1016/0893-6080(89)90020-8}}.

\bibitem{Funahashi1989}
K.-I. Funahashi, On the approximate realization of continuous mappings by neural networks, Neural Networks 2~(3) (1989) 183--192.
\newblock \href {https://doi.org/10.1016/0893-6080(89)90003-8} {\path{doi:10.1016/0893-6080(89)90003-8}}.

\bibitem{Leshno1993}
M.~Leshno, V.~Y. Lin, A.~Pinkus, S.~Schocken, Multilayer feedforward networks with a nonpolynomial activation function can approximate any function, Neural Networks 6~(6) (1993) 861--867.
\newblock \href {https://doi.org/10.1016/S0893-6080(05)80131-5} {\path{doi:10.1016/S0893-6080(05)80131-5}}.

\bibitem{Kingma:2014}
D.~P. Kingma, J.~Ba, {Adam: A Method for Stochastic Optimization}\href {http://arxiv.org/abs/1412.6980} {\path{arXiv:1412.6980}}.

\bibitem{rahaman2019spectral}
N.~Rahaman, A.~Baratin, D.~Arpit, F.~Draxler, M.~Lin, F.~A. Hamprecht, Y.~Bengio, A.~Courville, {On the Spectral Bias of Neural Networks} 97 (2019) 5301--5310.
\newblock \href {http://arxiv.org/abs/1806.08734} {\path{arXiv:1806.08734}}.

\bibitem{Xu:2019frequency}
Z.-Q.~J. Xu, Y.~Zhang, T.~Luo, Y.~Xiao, Z.~Ma, {Frequency Principle: Fourier Analysis Sheds Light on Deep Neural Networks}, Communications in Computational Physics 28 (2020) 1746--1767.
\newblock \href {http://arxiv.org/abs/1901.06523} {\path{arXiv:1901.06523}}, \href {https://doi.org/10.4208/cicp.OA-2020-0085} {\path{doi:10.4208/cicp.OA-2020-0085}}.

\bibitem{Jacot:2018ntk}
A.~Jacot, F.~Gabriel, C.~Hongler, {Neural Tangent Kernel: Convergence and Generalization in Neural Networks} (2018) 8571--8580\href {http://arxiv.org/abs/1806.07572} {\path{arXiv:1806.07572}}.

\bibitem{bietti2019inductive}
A.~Bietti, J.~Mairal, {On the Inductive Bias of Neural Tangent Kernels} (2019) 12897--12908\href {http://arxiv.org/abs/1905.12173} {\path{arXiv:1905.12173}}.

\bibitem{gunasekar2018characterizing}
S.~Gunasekar, J.~Lee, D.~Soudry, N.~Srebro, Characterizing implicit bias in terms of optimization geometry, in: Proceedings of the 35th International Conference on Machine Learning, Vol.~80 of Proceedings of Machine Learning Research, PMLR, 2018, pp. 1832--1841.
\newblock \href {http://arxiv.org/abs/1802.08246} {\path{arXiv:1802.08246}}.

\bibitem{arora2019fine}
S.~Arora, S.~S. Du, W.~Hu, Z.~Li, R.~Wang, Fine-grained analysis of optimization and generalization for overparameterized two-layer neural networks, in: Proceedings of the 36th International Conference on Machine Learning, Vol.~97 of Proceedings of Machine Learning Research, PMLR, 2019, pp. 322--332.
\newblock \href {http://arxiv.org/abs/1901.08584} {\path{arXiv:1901.08584}}.

\bibitem{CREATE}
{\relax King's College London}, \href{https://doi.org/10.18742/rnvf-m076}{{King's Computational Research, Engineering and Technology Environment (CREATE)}} (2022).
\newline\urlprefix\url{https://doi.org/10.18742/rnvf-m076}

\bibitem{Kos:2014bka}
F.~Kos, D.~Poland, D.~Simmons-Duffin, {Bootstrapping the $O(N)$ vector models}, JHEP 06 (2014) 091.
\newblock \href {http://arxiv.org/abs/1307.6856} {\path{arXiv:1307.6856}}, \href {https://doi.org/10.1007/JHEP06(2014)091} {\path{doi:10.1007/JHEP06(2014)091}}.

\bibitem{Kos:2016ysd}
F.~Kos, D.~Poland, D.~Simmons-Duffin, A.~Vichi, {Precision Islands in the Ising and $O(N)$ Models}, JHEP 08 (2016) 036.
\newblock \href {http://arxiv.org/abs/1603.04436} {\path{arXiv:1603.04436}}, \href {https://doi.org/10.1007/JHEP08(2016)036} {\path{doi:10.1007/JHEP08(2016)036}}.

\bibitem{Simmons-Duffin:2016wlq}
D.~Simmons-Duffin, {The Lightcone Bootstrap and the Spectrum of the 3d Ising CFT}, JHEP 03 (2017) 086.
\newblock \href {http://arxiv.org/abs/1612.08471} {\path{arXiv:1612.08471}}, \href {https://doi.org/10.1007/JHEP03(2017)086} {\path{doi:10.1007/JHEP03(2017)086}}.

\bibitem{Chang:2024whx}
C.-H. Chang, V.~Dommes, R.~S. Erramilli, A.~Homrich, P.~Kravchuk, A.~Liu, M.~S. Mitchell, D.~Poland, D.~Simmons-Duffin, {Bootstrapping the 3d Ising stress tensor}, JHEP 03 (2025) 136.
\newblock \href {http://arxiv.org/abs/2411.15300} {\path{arXiv:2411.15300}}, \href {https://doi.org/10.1007/JHEP03(2025)136} {\path{doi:10.1007/JHEP03(2025)136}}.

\bibitem{Zhu:2022gjc}
W.~Zhu, C.~Han, E.~Huffman, J.~S. Hofmann, Y.-C. He, {Uncovering Conformal Symmetry in the 3D Ising Transition: State-Operator Correspondence from a Quantum Fuzzy Sphere Regularization}, Phys. Rev. X 13~(2) (2023) 021009.
\newblock \href {http://arxiv.org/abs/2210.13482} {\path{arXiv:2210.13482}}, \href {https://doi.org/10.1103/PhysRevX.13.021009} {\path{doi:10.1103/PhysRevX.13.021009}}.

\bibitem{Han:2023yyb}
C.~Han, L.~Hu, W.~Zhu, Y.-C. He, {Conformal four-point correlators of the three-dimensional Ising transition via the quantum fuzzy sphere}, Phys. Rev. B 108~(23) (2023) 235123.
\newblock \href {http://arxiv.org/abs/2306.04681} {\path{arXiv:2306.04681}}, \href {https://doi.org/10.1103/PhysRevB.108.235123} {\path{doi:10.1103/PhysRevB.108.235123}}.

\bibitem{Hu:2023xak}
L.~Hu, Y.-C. He, W.~Zhu, {Operator Product Expansion Coefficients of the 3D Ising Criticality via Quantum Fuzzy Spheres}, Phys. Rev. Lett. 131~(3) (2023) 031601.
\newblock \href {http://arxiv.org/abs/2303.08844} {\path{arXiv:2303.08844}}, \href {https://doi.org/10.1103/PhysRevLett.131.031601} {\path{doi:10.1103/PhysRevLett.131.031601}}.

\bibitem{Halverson:2020trp}
J.~Halverson, A.~Maiti, K.~Stoner, {Neural Networks and Quantum Field Theory}, Mach. Learn. Sci. Tech. 2~(3) (2021) 035002.
\newblock \href {http://arxiv.org/abs/2008.08601} {\path{arXiv:2008.08601}}, \href {https://doi.org/10.1088/2632-2153/abeca3} {\path{doi:10.1088/2632-2153/abeca3}}.

\bibitem{Demirtas:2023fir}
M.~Demirtas, J.~Halverson, A.~Maiti, M.~D. Schwartz, K.~Stoner, {Neural network field theories: non-Gaussianity, actions, and locality}, Mach. Learn. Sci. Tech. 5~(1) (2024) 015002.
\newblock \href {http://arxiv.org/abs/2307.03223} {\path{arXiv:2307.03223}}, \href {https://doi.org/10.1088/2632-2153/ad17d3} {\path{doi:10.1088/2632-2153/ad17d3}}.

\bibitem{Halverson:2024axc}
J.~Halverson, J.~Naskar, J.~Tian, {Conformal fields from neural networks}, JHEP 10 (2025) 039.
\newblock \href {http://arxiv.org/abs/2409.12222} {\path{arXiv:2409.12222}}, \href {https://doi.org/10.1007/JHEP10(2025)039} {\path{doi:10.1007/JHEP10(2025)039}}.

\bibitem{Ferko:2026axm}
C.~Ferko, J.~Halverson, A.~Mutchler, {Universality of Neural Network Field Theory} (1 2026).
\newblock \href {http://arxiv.org/abs/2601.14453} {\path{arXiv:2601.14453}}.

\bibitem{Mazac:2018ycv}
D.~Mazac, M.~F. Paulos, {The analytic functional bootstrap. Part II. Natural bases for the crossing equation}, JHEP 02 (2019) 163.
\newblock \href {http://arxiv.org/abs/1811.10646} {\path{arXiv:1811.10646}}, \href {https://doi.org/10.1007/JHEP02(2019)163} {\path{doi:10.1007/JHEP02(2019)163}}.

\bibitem{Liendo:2012hy}
P.~Liendo, L.~Rastelli, B.~C. van Rees, {The Bootstrap Program for Boundary CFT$_d$}, JHEP 07 (2013) 113.
\newblock \href {http://arxiv.org/abs/1210.4258} {\path{arXiv:1210.4258}}, \href {https://doi.org/10.1007/JHEP07(2013)113} {\path{doi:10.1007/JHEP07(2013)113}}.

\bibitem{Dolan:2006ec}
F.~A. Dolan, M.~Nirschl, H.~Osborn, {Conjectures for large N superconformal N=4 chiral primary four point functions}, Nucl. Phys. B 749 (2006) 109--152.
\newblock \href {http://arxiv.org/abs/hep-th/0601148} {\path{arXiv:hep-th/0601148}}, \href {https://doi.org/10.1016/j.nuclphysb.2006.05.009} {\path{doi:10.1016/j.nuclphysb.2006.05.009}}.

\end{thebibliography}

\end{document}